\documentclass[aps,12pt,preprint,superscriptaddress,showpacs]{revtex4}
\usepackage{psfig}
\usepackage{longtable}
\begin{document}
\title{On the absence of appreciable half-life changes in alpha emitters cooled in metals to 1 Kelvin and below.}
\author{N.~J.~Stone} 
\affiliation{Department of Physics,
University of Oxford, OX1 3PU Oxford, UK}
\affiliation{Department of Physics and Astronomy, University of Tennessee, Knoxville, TN 37996, USA}
\author{J.~R.~Stone} 
\affiliation{Department of Physics,
University of Oxford, OX1 3PU Oxford, UK}
\affiliation{Department of Chemistry and Biochemistry, University of Maryland, College Park, MD 20742, USA}
\author{M.~Lindroos}
\altaffiliation[Present address:] { AB Department, CERN, 1211 Geneva 23, Switzerland}
\affiliation{Department of Physics,
University of Oxford, OX1 3PU Oxford, UK}
\author{P.~Richards}
\altaffiliation [Present address:] { Information and Learning Services, University of Plymouth, PL4 8AA Plymouth, UK}
\affiliation{Department of Physics,
University of Oxford, OX1 3PU Oxford, UK}
\author{M.~Veskovic} 
\altaffiliation [Present address:] { Department of Physics, University of Novi Sad, 21 000 Novi Sad, Serbia}
\affiliation{Department of Physics,
University of Oxford, OX1 3PU Oxford, UK}
\author{D.~A.~Williams}
\altaffiliation [Present address:] { Sungard, 3 Milton Park, OX14 4RN Abingdon, UK}
\affiliation{Department of Physics,
University of Oxford, OX1 3PU Oxford, UK}
\date{\today}
\begin{abstract}
The recent suggestion that dramatic changes may occur in the lifetime of alpha and beta decay when the activity, in a pure metal host, is cooled to a few Kelvin, is examined in the light of published low temperature nuclear orientation (LTNO) experiments on such sources cooled to as low as 25 mK, with emphasis here on alpha decay. In LTNO observations are made of the anisotropy of radioactive emissions with respect to an axis of orientation. Correction of data for decay of activities in metallic samples held at temperatures at and below 1 Kelvin for periods of several days has been a routine element of LTNO experiments for many years. No evidence for any change of half-life on cooling, with an upper level of order 1\%, has been found, in striking contrast to the predicted changes, for alpha decay, of several orders of magnitude. The proposal that such dramatic changes might alleviate problems of disposal of long-lived radioactive waste is shown to be unrealistic.
\end{abstract}

\maketitle
\section{\label{sec:intro}Introduction}
In a recent paper ideas were presented which led to the suggestion that the half-lives of alpha emitters, particularly those of lower decay energy, would be strikingly reduced if the activity was in a metallic environment and cooled to temperatures of a few Kelvin as compared to `normal' decay in insulating materials at room temperature \cite{ket06}. The authors surmised that, through the influence of the conduction, `valence' or `free', electron potential near the nucleus, the alpha decay barrier would be effectively lowered and thus cooling of long-lived radioactive waste might lead to their much more rapid decay and safe disposal. No experimental evidence was offered in support of these suggestions.
It is the objective of this paper to report aspects of already published experiments of significance to these suggestions, clearly indicating that the predictions are far from experimental reality. The experiments concern the decay of relatively long lived $^{\rm 224}$Rn, $^{\rm 225}$Ra and $^{\rm 227}$Ac and their daughter decay chains implanted in an iron metal foil and studied at temperatures down to below 20 mK. The original objectives of this work were to study the polarization of the decaying isotopes through measurement of the angular properties of their alpha and gamma emissions - the method of low temperature nuclear orientation (LTNO) - leading to results on the hyperfine interactions of the isotopes involved, their magnetic dipole moments and the magnetic hyperfine fields they experienced as implants in the ferromagnetic iron lattice \cite{lin92a,lin92b,wil96}. However the results throw direct light upon the suggestions made by Kettner et al in Ref.~\cite{ket06}. Brief mention of evidence from experiments involving beta decay is made in the discussion.

This paper presents experimental evidence only, without detailed comment on the theoretical considerations on which the proposals of Ref.~\cite{ket06} were based. We note however that (i) serious objections to the underlying ideas on the effect of screening on alpha particle Coulomb barrier penetration have been raised elsewhere \cite{zin06} and (ii) that the suggestion that cooling to low temperatures might be worthwhile depends upon treating conduction electrons as classical gas with mean energy proportional to temperature T. Furthermore, the idea that screening can take place by electrons at a Debye radius much smaller than the K-shell mean radius \cite{ket06,jep07}, falling towards nuclear radii at and below 1K, contravenes basic quantum mechanics. There are simply no states available to construct such confined electron wavefunctions. If there were they would be occupied by the atomic electrons. More comments are given in the discussion.

The experiments which provide the evidence given here were performed at ISOLDE, CERN and Oxford over a period of several years. Other LTNO groups, in particular those from Leuven and Bonn, have published work on implanted alpha emitters cooled to millikelvin temperatures in metals, containing similar evidence (see e.g. \cite{wou91,sch99,sev05} and references therein).

The basic ideas of the proposed enhancement mechanism are given in Sec.~\ref{sec:basid}, which is followed in Sec.~\ref{sec:ltno} by a survey of the ways in which low temperature nuclear orientation experiments are sensitive to source lifetime and decay correction. Sec.~\ref{sec:impl} gives details of the preparation of the implanted metallic samples used in the work referred to and Sec.~\ref{sec:pred} introduces model calculations based on three possible scenarios regarding the correct lifetimes to use in analysis of the experiments. Sec.~\ref{sec:expbeh} gives an extended analysis of predictions and experimental results for the three decay chains headed by $^{\rm 224}$Rn, $^{\rm 225}$Ra and $^{\rm 227}$Ac. Discussion and conclusions are presented in Sec.~\ref{sec:discon}.

\section{\label{sec:basid}Basic ideas of the proposed alpha decay enhancement mechanism}
From Ref.~\cite{ket06} the enhancement of alpha decay rates in a metal is associated with the presence around the metallic ion of a screening charge, due to free electrons, which is taken to reduce the barrier height by the screening energy U$_{\rm D}$. According to this picture, the reduced barrier height enhances the rate of alpha tunneling through the barrier to separate from its parent nucleus, thus reducing the lifetime. Since ionic charge screening by free electrons occurs only in metals, it is predicted that large differences of alpha decay lifetime will be observed depending upon whether the decay takes place in an insulator or a metal.  An expression is given for U$_{\rm D}$  in terms of the charges of the alpha particle Z$_\alpha$ and the daughter nucleus Z$_{\rm t}$, the temperature of the lattice electrons and a parameter U$_{\rm e}$(d+d) which relates to a $d+d$ fusion reaction in a metal \cite{ket06}:
\begin{equation}
U_{\rm D} = Z_\alpha Z_{\rm t}U_{\rm e}(d+d)_{290} \left(\frac{290}{T}\right)^{\rm 1/2}
\label{eq1}  							
\end{equation}
where $T$ is the absolute temperature of the metal in which the decay takes place. A `typical' value of U$_{\rm e}$(d+d)$_{\rm 290}$ is given to be 300 eV at 290 Kelvin \cite{ket06}. The temperature dependence arises from taking the Drude model for the `quasi-free' valence electrons as having an (average) kinetic energy of 0.5 kT. Based on this expression it is suggested that
by cooling the activity in a metal sample to liquid helium temperature (4.2 K), the half-life of  $^{\rm 210}$Po can be reduced from 138 d to 0.5 d and of  $^{\rm 226}$Ra from 1600 y to 1.3 y, factors of order 1000, in each case through a reduction of the effective barrier potential by about 420 keV (see Ref.~\cite{ket06}).

The phenomenon of ionic charge screening in metals is well established and its consequence is giving rise to small energy shifts, of order several hundred electronvolts, for cross sections in low energy reaction studies in metallic targets, is fully documented in Ref.~\cite{ket06}  and other sources. The issue in this paper is to explore whether, through a strong dependence of the effect on temperature, charge screening can lead to dramatic consequences at temperatures of order a few degrees and below.
 Three scenarios are considered and their consequences compared with existing experimental evidence. They are:
\begin{itemize}
\item The Full effect: the suggested mechanism operates fully at all temperatures.
\item The Partial effect: an attenuated version of the mechanism occurs, leading to a 5\% reduction in all alpha decay half-lives at 1 K, which however obeys the predicted U$_{\rm D}$ temperature dependence.
\item The Null effect: all effects of screenimg on half-life are negligibly small so that `normal' half-lives may be used. 
\end{itemize}

\section{\label{sec:ltno}Low temperature nuclear orientation experiments and their sensitivity to half-lives.}
Low temperature nuclear orientation experiments involve cooling radioactive isotopes in samples in which they experience strong hyperfine interactions to temperatures at which they become, in the case of a magnetic interaction, polarised along the direction of the magnetic field. The appropriate temperature range is typically between 1 and 100 mK for activities in, for example, a ferromagnetic iron foil sample. For relatively long-lived activities such as are concerned here, the experiments are performed with the sample soldered to the cold finger of a dilution refrigerator and involve initial cooling, by stages, to liquid nitrogen (77 K) and liquid helium (4.2 K) temperatures, each for about 24 hours to allow for cooling and establishment of good vacuum conditions (less then 10$^{\rm -6}$ mbar), before reaching 1 K attached to a pumped liquid helium reservoir. At 1 K temperature the nuclei remain unoriented and `warm' counts, to be used for normalisation of the measured anisotropies at lower temperatures, are taken for several hours. Circulation of the $^{\rm3}$He/$^{\rm4}$He mixture in the refrigerator is started and after a few hours the sample temperature enters the millikelvin range. It can be kept there for many days, but often circulation is stopped to allow return to 1 K to take more `warm' counts in order that the decay of the sample can be fully allowed for in analysis.
In the orientation measurements themselves the basic observable is the anisotropy of angular distribution of emission of particles or radioation from an oriented source
\begin{equation}
\mathcal{A}=(W(\theta,T)-1)\%
\end{equation}
\noindent
where $W(\theta,T)$ is the angular distribution function at angle $\theta$ with respect to the axis of orientation and temperature $T$. It is derived form experimental count rates by
\begin{equation}
W_{\rm exp}(\theta,T)=\frac{N^{\rm (c)}}{N^{\rm (w)}}.
\end{equation}
\noindent
$N^{\rm (c)}$ and $N^{\rm (w)}$ are decay and dead-time corrected count rates in a detector when the source in `cold' (oriented at millikelvin temperature) and `warm' (unoriented at $\sim$ 1 K or above), respectively. The decay correction, fundamental to these measurements, is directly relevant to the present work, as it involves the decay constant of a isotope in question through a correction factor
\begin{equation}
\Xi=\frac{\lambda \delta \exp(+\lambda t)}{1-\exp(-\lambda \delta)}
\end{equation}
\noindent
where $\lambda$=ln(2)/T$_{\rm 1/2}$ is the decay constant, $\delta$ is the real time of the counting period and $t$ is the difference between the arbitrary normalisation time and the start of the measurement. After this correction is applied, each count rate can be compared to rates in spectra taken at different times as though they were taken at the same instant in time.
Physical parameters are extracted from measured anisotropies by comparison of the experimental angular distribution and the theoretical expression
\begin{equation}
W(\theta,T)=1+f\sum\limits_{k=2}^{k_{\rm max}}B_{\rm k}(T) U_{\rm k} A_{\rm k} Q_{\rm k} P_{\rm k}(cos\theta)
\end{equation}
\noindent
where B$_{\rm k}$, U$_{\rm k}$ and A$_{\rm k}$ are the usual orientation, deorientation and angular distribution coefficients, Q$_{\rm k}$ are the detector solid angle corrections, P$_{\rm k}$ are Legendre polynomial and $f$ is the fraction in good sites \cite{sto86}. Depending on the nature of a particular experiment, selected information contained in one of the coefficients can be obtained by such a comparison, assuming all the other coefficients are known. In the work referred to in this paper, the orientation coefficients B$_{\rm k}$, dependent on the product of magnetic dipole moment and the hyperfine field acting on the studied nuclei in the sample, were used to extract the magnetic dipole moment of oriented states and compare them with either the same moments obtained by a different technique or with systematics and theory. It is obvious that should the experimental anisotropy be determined incorrectly, i.e. if the correction for the radioactive decay, made by using decay constants known at room temperature, would not apply at temperatures of 1 K and lower, the extracted magnetic dipole moment would be wrong.

Additional sensitivity to the appropriateness of the decay correction arises from other detailed properties of the measured anisotropic distributions. The first sensitive feature is that, at the lowest temperatures, the anisotropy must become independent of temperature as the nuclei become fully polarized. Since there is a time sequence in such measurements as the sample temperature is changed, any incorrectness in the source decay correction will impose a false slope or scatter upon the extracted temperature dependence. The second sensitive feature is that the anisotropy involves an increase in observed intensity in some directions to the orientation axis, and in others a decrease, the actual quantitative changes being determined by well established theory of the distribution. The application of an incorrect adjustment for source decay will falsify the relationship between observations in detectors in different directions to the axis. Thus, if the decay correction assumes too short a lifetime, the `warm' unoriented distribution will be taken to fall too fast, so that any observed increase in intensity, measured at a time later than the `warm' distribution, will be analysed as too large relative to its true value by a certain factor, and any simultaneously observed decrease would be analysed as smaller than its true value, by the same factor. These failures would lead to distortion of the relation between the analysed anisotropies in the two directions, which can usually be readily identified.

A separate aspect of the LTNO technique with relevance to lifetime is the fact that for the observed anisotropies to be characteristic of the hyperfine interaction in a specific nuclear state, usually a ground state, the lifetime of that state must be long enough for the nuclei to reach thermal equilibrium with the metallic lattice and achieve a degree of polarisation characteristic of that temperature and interaction strength. The time for this to be reached, the nuclear spin-lattice relaxation time, T$_{\rm 1}$, has been established in implanted, diffused and co-melted metallic samples as following the Korringa conduction electron mechanism. In iron samples of the type discussed in this paper T$_{\rm 1}$'s have been measured to range from many hours to below 100 ms at millikelvin temperatures. The relevant parameter is the ratio  T$_{\rm 1}/$ T$_{\rm 1/2}$. There is abundant evidence that were a lifetime to become shorter than 0.1 ms thermal equilibrium will not be achieved. In situations involving a decay cascade, such as are described in this paper, a degree of polarisation may be 'inherited' from a precursor state higher in the decay chain. Reference to situations of this kind are made in the text where relevant, as the degree of equilibration will be affected if lifetimes were to be severely reduced on cooling.

Finally, as a matter of course, OLNO experiments include measurement of the unoriented intensity at both the start and the end of the experiment, and quite often also at an intermediate time, when the temperature is at close to 1 K. It is a standard requirement that these normalisation measurements must, after due correction for decay of the source, yield a consistent value for the reference `warm' intensity to be used in evaluation of all measured anisotropies. If the normalisation should not be consistent this would be immediately apparent and present a challenge to the experimenters. Thus we have, in effect, a measure of the source decay at 1 K over the period of the experiment, which is typically at least several days.

Each experiment has individual sensitivity to sample half-life and the cases of the three alpha decay chains described here are discussed in detail in Sec.~\ref{sec:expbeh}.

\section{\label{sec:impl}Implanted metallic sample preparation for LTNO alpha anisotropy experiments.}
For the experiments described here, alpha active samples of $^{\rm 224}$Ra, $^{\rm 225}$Ra and $^{\rm 227}$Ac were prepared at the ISOLDE isotope separator facility, CERN, the parent activities being implanted at 60 keV into prepared 99.999\% pure iron foils which have been carefully surface polished. The implant dose was $\sim$2$\cdot$10$^{\rm 13}$ for mass A= 225 and A=227 and $\sim$2$\cdot$10$^{\rm 11}$ for mass A=224. Experience of implantation of a wide range of elements has shown that the implantation energy, 60 keV, is amply sufficient to allow the implanted ions to penetrate the omnipresent thin oxide layer on the foil surface and come to rest in undisturbed pure iron. Estimates based on calculated projected ranges \cite{range} show that an ion with a nucleus of mass A$\sim$200 accelerated through 60 keV will penetrate $\sim$14 nanometers, or about 75 atomic layers in iron. Allowing for an oxide layer of approximately 5 nanometers \cite{her86} the average depth of the initially implanted alpha emitters is $\sim$45 atomic layers of pure iron. In samples where there is an alpha decay cascade, recoil following alpha emission alters the initial depth distribution. The $\sim$100 keV nuclear recoil energy can cause ejection of the daughter from the sample, but, on average, the mean implant depth of those nuclei remaining in the pure iron, about 75\% of the initial quantity after a three-stage cascade, is increased considerably compared with the initial depth \cite{Wthesis}. Many experiments have demonstrated that this implantation energy is adequate to describe the environment of the implants as fully typical of the metal. In the context of low temperature nuclear orientation evidence comes from the fact that Nuclear Magnetic Resonance (NMR) of the implants has been observed at the same frequencies in implanted samples as in samples prepared by thermal diffusion to depths of $\sim$1 micron \cite{her77}. Additional evidence of the implants being in true metallic lattice is the fact that the implanted nuclei have been observed to relax in on-line implantation experiments with time constants  T$_{\rm 1}$ described fully by the Korringa conduction electron mechanism \cite{sha89}.

\section{\label{sec:pred}Predictions of alpha activity changes on cooling}
As mentioned is Sec.~\ref{sec:basid} we consider detailed estimates of effects of proposed lifetime reductions in three possible scenarios to be referred to as the Full, Partial, and Null cases. In all cases the lifetime reduction estimates, based on the expression for the potential U$_{\rm D}$, as given in Ref.~\cite{ket06} and quoted above, can be readily estimated, without detailed calculation, by taking the systematic half-lives of allowed alpha emitters versus decay energy (see e.g. \cite{kra87}) to obtain d(log T$_{\rm 1/2}$)/dE$_\alpha $ as a function of the alpha particle decay energy, E$_\alpha$.
  
\begingroup 
\squeezetable 
\begin{longtable*}{cccccccc} 
 \caption{\label{tab1}Predicted alpha decay lifetime reductions and alpha fraction changes based upon Ref.~\cite{ket06}}\\
\hline \hline
\multicolumn{1}{c}%
\textsc{Parent}  &   \textsc{Normal alpha} & \textsc{Alpha}  & \textsc{Normal} 
&  \textsc{U$_{\rm D}$} & \textsc{half-life} & \textsc{predicted alpha}& \textsc{predicted}\\ 
\textsc{Nucleus}& \textsc{decay half-life}  &   \textsc{Energy} & \textsc{fraction} 
& \textsc{(keV)} & \textsc{reduction} &\textsc{fraction}  &
\textsc{decay halflife} \\ \textsc{} & 
\textsc{at 290 K *} & \textsc{ (keV)}& \textsc{(\%) at 290 K} & \textsc{ } &
\textsc{factor at 1K} & \textsc{(\%) at 1K **} & \textsc{(\%) at 1K ***}
 \\ 
\hline 
\endfirsthead
\caption[]{(continued)}\\
\hline \hline
\multicolumn{1}{c}%
\textsc{Parent}  &   \textsc{Normal alpha} & \textsc{Alpha}  & \textsc{Normal} 
&  \textsc{U$_{\rm D}$} & \textsc{halflife} & \textsc{predicted alpha}& \textsc{predicted}\\ 
 \textsc{Nucleus}& \textsc{decay halflife}  &   \textsc{Energy} &  
\textsc{fraction}  & \textsc{(keV)} & \textsc{reduction} & \textsc{fraction} &
\textsc{decay halflife} \\ \textsc{} & 
\textsc{at 290 K *} & \textsc{ (keV)}& \textsc{(\%) at 290 K} & \textsc{ } &
\textsc{factor at 1K} & \textsc{(\%) at 1K **} & \textsc{(\%) at 1K ***}
 \\ 
\hline  
\endhead
\\
\multicolumn{8}{l}{A=224 decay chain, headed by alpha emitter $^{\rm 224}$Ra}   \\    
$^{\rm 224}$Ra	&3.66 d	      &5686  &94     &898   & 6.3 x 10$^{\rm 4}$   &        & 4.7 s \\	
		&             &5449  &5.5    &898   & 1.3 x 10$^{\rm 5}$	\\ \\
$^{\rm 220}$Rn	&55.6 s	      &6228  &100    &878   & 1.2 x 10$^{\rm 4}$   &        & 4.7 ms \\ \\
$^{\rm 216}$Po	&0.15 s       &6779  &100    &857   & 2.7 x 10$^{\rm 3}$   &        &55 x 10$^{\rm -6}$ s \\ \\
$^{\rm 212}$Bi	&60.6 m       &6090  &27.2   &847   & 1.2 x 10$^{\rm 4}$   &26      &0.82 s \\
                &             &6051  &69.9   &847   & 1.3 x 10$^{\rm 4}$   &74      \\ \\
\multicolumn{8}{l}{predicted missing transitions if enhanced alpha decay of $^{\rm 212}$Bi weakens beta decay branch}\\
$^{\rm 212}$Po           &0.30 x 10$^{\rm -6}$ &8784  &100 &\multicolumn{3}{l}{not seen if beta decay of $^{\rm 212}$Bi ceases}&  \\ \hline \\
\multicolumn{8}{l}{A=225 decay chain, headed by $^{\rm 225}$Ra $>$99\% beta $<$1 x 10$^{\rm -4}$\% alpha T$_{\rm 1/2}$ 14.8 d} \\
$^{\rm 225}$Ac  &10.0 d      &5830   &50.7   &889   &3.8 x 10$^{\rm 4}$   &46       &20.6 s \\	
                &            &5794   &24.3   &889   &4.3 x 10$^{\rm 4}$   &25  \\	
                &            &5732   &10.1   &889   &5.0 x 10$^{\rm 4}$   &12  \\
                &            &5637   &4.4    &889   &6.5 x 10$^{\rm 4}$   &7   \\ \\
$^{\rm 221}$Fr  &4.8 m       &6341   &83.4   &867   &7.7 x 10$^{\rm 3}$   &77       &34 ms \\	
                &            &6126   &15.1   &867   &1.3 x 10$^{\rm 4}$   &23   \\ \\	
$^{\rm 217}$At	&32.3 ms     &7067   &99.9   &867   &1.8 x 10$^{\rm 3}$	  &100      &18 x 10$^{\rm -6}$ s \\ \\
$^{\rm 213}$Po	&4 x 10$^{\rm -6}$ s &8375 &99.9 &857 &3.3 x 10$^{\rm 2}$ &100      &1.2 x 10$^{\rm -8}$ s \\ \\
\multicolumn{8}{l}{predicted additional transitions via enhanced alpha decay of $^{\rm 225}$Ra} \\ 
$^{\rm 225}$Ra  &$>$4 x 10$^{\rm 4}$y &4950  & not seen &898   &5.0 x 10$^{\rm 5}$   &30\% of all     &28 d\\ \\
$^{\rm 221}$Rn  &125 m        &6035  & strongest &878  &2.0 x 10$^{\rm 4}$    &strongest & 0.38 s \\ \\ 
$^{\rm 217}$Po  &$<$10s       &6537  & $>$95     &857  &4.4 x 10$^{\rm 3}$  &100       &$<$ 2 ms \\ \\
\multicolumn{8}{l}{predicted additional transitions via enhanced alpha decay of $^{\rm 213}$Bi}\\
$^{\rm 213}$Bi           & 38 h        &5869  & 93    &847   & 2.1 x 10$^{\rm 4}$   & 85     & 5.9s \\
                &             &5549  &  7    &847   & 4.8 x 10$^{\rm 4}$   & 15 \\ \hline \\
\multicolumn{8}{l}{A=227 decay chain headed by $^{\rm 227}$Ac , normal decay; 98.6\% beta, 1.4\% alpha, T$_{\rm 1/2}$ 21.8 y}\\ 
$^{\rm 227}$Th	&  18.72 d   &6038   &24.5   &918   &3.2 x 10$^{\rm 4}$   &19.7     &40.7 s   \\
                &100\% alpha &5979   &23.4   &918   &3.5 x 10$^{\rm 4}$   &20.5 \\
		&            &5756   &20.3   &918   &6.8 x 10$^{\rm 4}$   &34.7 \\
		&            &5713   &4.9    &918   &7.6 x 10$^{\rm 4}$   &9.3  \\
		&            &5709   &8.2    &918   &7.6 x 10$^{\rm 4}$   &15.6 \\ \\
$^{\rm 223}$Ra	&  11.44 d   &5748   &9.1    &898   &5.4 x 10$^{\rm 4}$	  &7        &14.0 s   \\
                &100\% alpha &5717   &53.7   &898   &6.1 x 10$^{\rm 4}$   &46.7 \\
                &            &5607   &26     &898   &7.9 x 10$^{\rm 4}$	  &29.1 \\
                &            &5539   &9.1    &898   &9.9 x 10$^{\rm 4}$   &12.7 \\
                &            &5434   &2.3    &898   &13.5 x 10$^{\rm 4}$  &4.4  \\  \\
$^{\rm 219}$Rn	&   4 s      &6819   &81     &878   &3.1 x 10$^{\rm 3}$	  &62.9     &1.1 ms   \\
                &100\% alpha &6553   &11.5   &878   &5.3 x 10$^{\rm 3}$	  &16.6 \\
                &            &6425   &7.5    &878   &7.2 x 10$^{\rm 3}$	  &14.7 \\  \\
$^{\rm 215}$Po  &1.8 ms      &7386   &100    &857   &1.0 x 10$^{\rm 3}$   &100      &1.8 x 10$^{\rm -6}$ s  \\ \\
$^{\rm 211}$Bi  &2.15 m	     &6623   &84     &847   &3.3 x 10$^{\rm 3}$	  &71         &33 ms  \\
                &            &6279   &16     &847   &7.1 x 10$^{\rm 3}$	  &29   \\  \\
\multicolumn{8}{l}{predicted additional transitions via enhanced alpha decay of $^{\rm 227}$Ac} \\			
$^{\rm 227}$Ac	&21.77 y     &4950   &47     &889   &5.7 x 10$^{\rm 5}$   &45.4     &1.00 d	\\
                &            &4938   &40     &889   &5.7 x 10$^{\rm 5}$   &38.8 \\	
                &            &4870   &6.1    &889   &7.5 x 10$^{\rm 5}$   &7.8  \\	
                &            &4853   &3.7    &889   &7.5 x 10$^{\rm 5}$   &4.7  \\  \\	
$^{\rm 223}$Fr	&21.8 m	     &5340   &strongest  &867   &1.3 x 10$^{\rm 5}$	  &100      &3.4 m \\ \\
$^{\rm 219}$At	&0.9 m       &6275   &strongest  &867   &4.0 x 10$^{\rm 3}$	  &100      &13.5 ms \\ \\ 
 \hline
\multicolumn{5}{l}{* allowing for beta decay branching}\\					
\multicolumn{5}{l}{** after renormalisation of alpha branching}\\
\multicolumn{5}{l}{***neglecting any change in beta half-life}\\	
\hline \hline				
\end{longtable*}
\endgroup
\noindent

This simple approach is valid as the half-life adjustment relates to the barrier penetration factor, not to the pre-formation of the alpha particle within the parent nucleus. This method, with estimates of U$_{\rm
D}$ direct from Eqn.~(\ref{eq1}) and the typical room temperature value of U$_{\rm e}$(d+d)$_{\rm 290}$ given in Ref.~\cite{ket06}, has been used to obtain estimated half-life reductions, see Table~\ref{tab1}, for the principal alpha decay transitions involved in the three isotope chains relevant to this paper at 1 K. These are examples of the Full scenario. Estimates of changes at different temperatures (77 K, 4.2 K and 25 mK) for the Full scenario have been made by the same method. For the Partial scenario it is assumed that the lifetime change is -5\% at 1 K and the simple theory have been used to give the value of U$_{\rm D}$ to which this reduced effect would correspond.
It has been then further assumed that, to estimate the effects at different temperatures, the temperature dependence follows the predicted T$^{\rm -1/2}$ relationship of Eqn.~\ref{eq1}. The Null scenario makes no allowance for any half-life change with temperature. This is the expectation were the screening electrons to obey Fermi-Dirac statistics and has been used implicitly in all published OLNO experiments.
\begin{figure}
\vspace*{0.3cm}
\centerline{\psfig{file=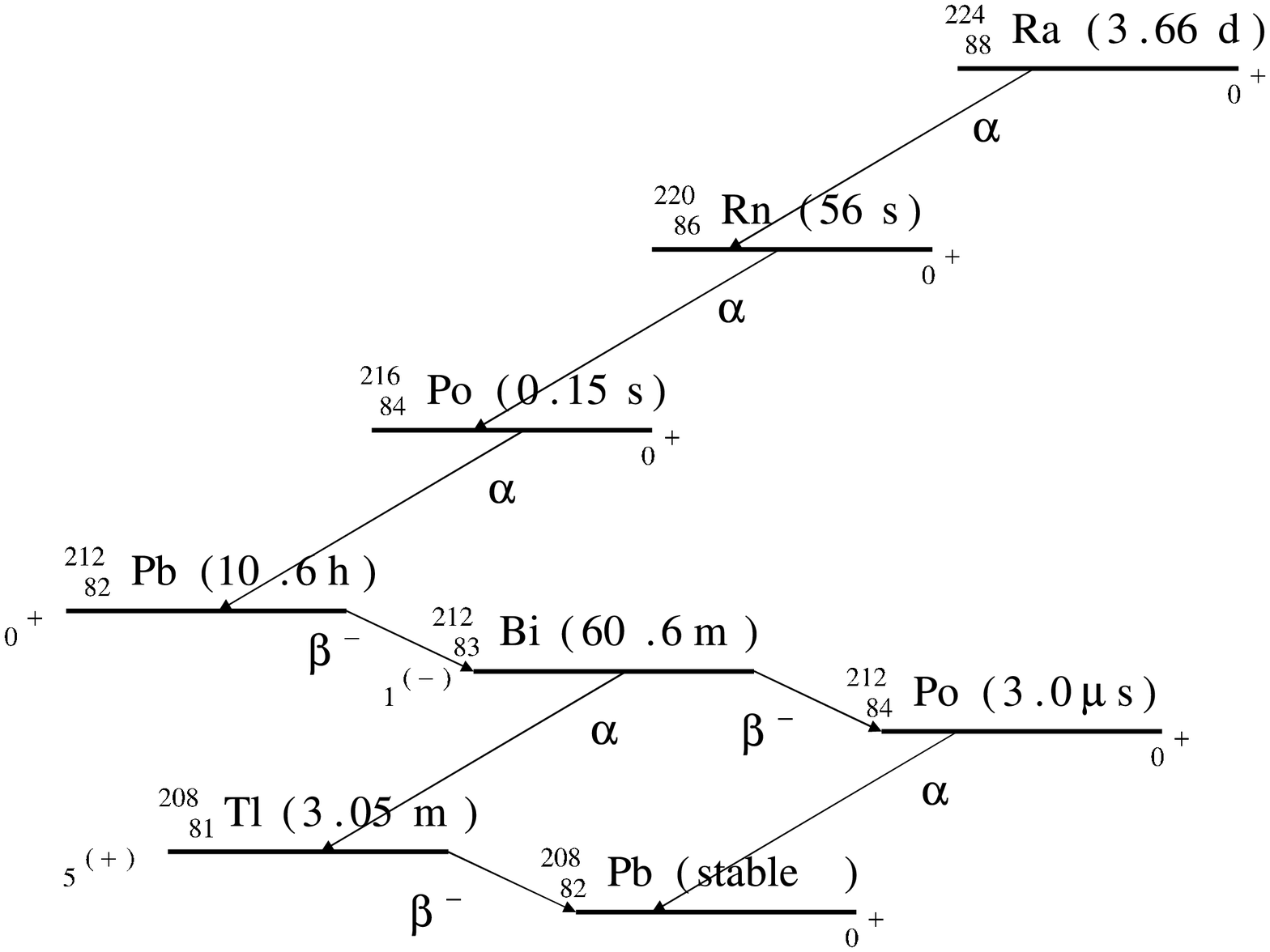,height=8cm,width=10cm}}
\caption{\label{fig1}$^{\rm 224}$Ra decay chain; $\beta^-1$/$\alpha$ branching ratio from $^{\rm 212}$Bi predicted to weaken.}
\end{figure}
\section{\label{sec:expbeh}Experimental behaviour of cooled alpha sources implanted into iron, compared with the Full, Partial and Null scenarios}

The decay chains of the three implanted isotopes listed above are shown in Figs.~\ref{fig1},\ref{fig5} and \ref{fig8} and details of the lifetimes and stronger alpha decay transitions in each decay are given in Table~\ref{tab1}. All half-lives are taken from \cite{toi}. The consequences of the predicted half-life changes are qualitatively different for the three chains as outlined below. 

\subsection{The $^{\rm 224}$Ra decay chain.}
\begin{figure}
\vspace*{0.3cm}
\centerline{\psfig {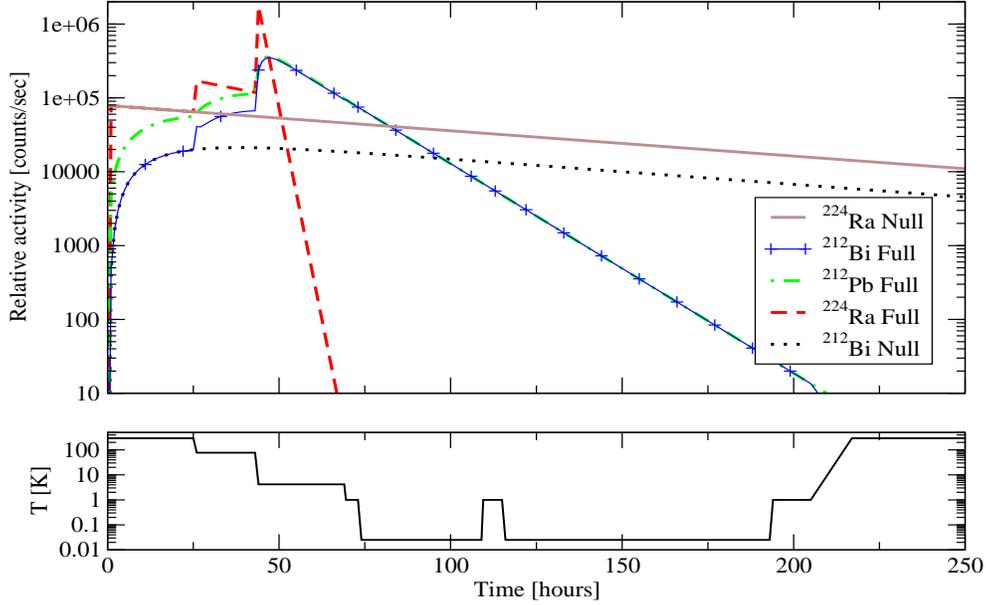}}
\caption{\label{fig2}Relative activity for selected members of the $^{\rm 224}$Ra decay chain in the Full and Null scenarios (top panel) and sample temperature (bottom panel) as a function of time after implantation.}
\end{figure}

The implanted isotope, $^{\rm 224}$Ra is an alpha emitter (3.66 d) and a single sample was studied during an experiment at the NICOLE dilution refrigerator at ISOLDE, CERN, which lasted for about 8 days starting one day after the sample was implanted. The time and temperature sequence of this experiment is shown in Fig.~\ref{fig2} which also illustrates the predictions based on the Full and Null scenarios.
The Null scenario is simple in that, after the first 20 hours or so, all activities are predicted to follow the 3.66 d half-life of the long-lived parent. The Full scenario, at the other extreme, predicts that the parent activity, which starts out as the upper line in Fig.~\ref{fig2}, should show two sharp step increases as the sample is cooled, first to liquid nitrogen temperature, 77 K, then to 4.2 K. At this temperature the predicted half-life is already shorter than the time which elapsed before being cooled further to 1 K, at which the lifetime is predicted to be 4.7 s (Table~\ref{tab1}) so the $^{224}$Ra activity falls away very steeply. The three prominent alpha transitions at  5686 keV ($^{\rm 224}$Ra), 6288 keV ($^{\rm 220}$Rn) and 6779 keV ($^{\rm 216}$Po) should thus essentially vanish before the OLNO experiment started, leaving only the (temporarily increased) $^{\rm 212}$Pb beta activity, and its daughters, present in the source, with alpha transitions at 6090 keV and 6051 keV, which decay with the normal $^{\rm 212}$Pb half-life, 10.6 h.

The LTNO experiment has two periods at very low temperature, taken to be for simplicity a steady 25 mK, with periods at 1K at the start, after about 110 hours and at the end, about 190 hours after the sample was made. It is clear that, in the Full scenario, the $^{212}$Bi activity should fall steadily throughout the experiment with the a 10.6 h half-life. 
\begin{figure}
\vspace*{0.3cm}
\centerline{\psfig{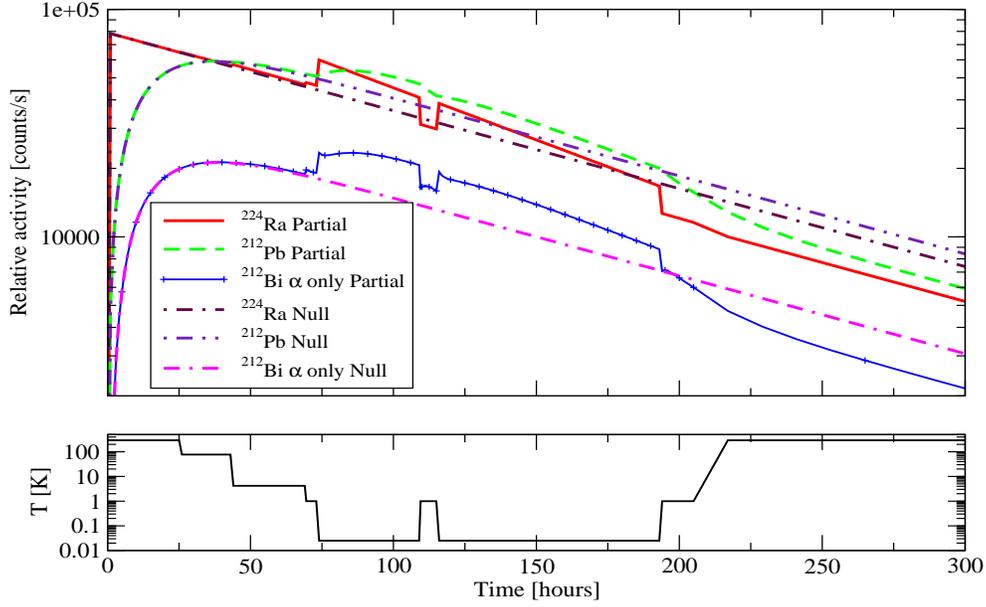}}
\caption{\label{fig3} As Fig.~\protect\ref{fig2} but the Partial and Null scenarios.}
\end{figure}

The Partial scenario is illustrated in Fig.~\ref{fig3}, which compares the predictions of this case with the Null scenario. For the Partial case there is little change until the sample reaches 1 K at which there is a small step upward in all decay rates. A much larger step occurs on cooling to 25 mK, immediately in the case of $^{\rm 224}$Ra and delayed through the (unchanged) 10.6 h beta decay half-life of $^{\rm 212}$Pb for the activities of $^{\rm 212}$Pb and $^{\rm 212}$Bi. The alpha decay fraction of  $^{\rm 212}$Bi is predicted to increase from its normal value of 36\% to 37.2\% at 1 K and 43.5\% at 25 mK. It is clear from Fig.~\ref{fig3} that a set of `warm' counts taken during the three periods of the experiment with the sample at 1 K would not follow the predictions of the Null scenario were the Partial scenario to be correct. Furthermore large distortions of the ratios of 25 mK activity to 1 K activity in detectors placed at angles for which the true anisotropy would lead to increase/decrease of counts rates would be produced by the common increases predicted on cooling the sample to 25 mK.

\subsubsection{Experimental findings for the $^{\rm 224}$Ra decay chain.}
\begin{figure}
\vspace*{0.3cm}
\centerline{\psfig{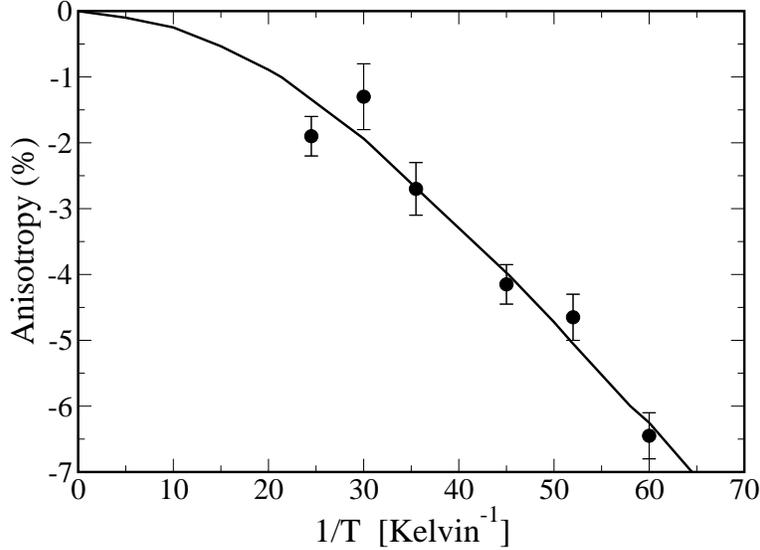}}
\caption{\label{fig4}$^{\rm 224}$Ra decay chain: $^{\rm 212}$Bi 727 keV gamma transition anisotropy vs inverse temperature \protect\cite{lin92a}.}
\end{figure}
In the published experiment on this decay chain only data the 727 keV gamma transition from the daughter of $^{\rm 212}$Pb, $^{\rm 212}$Bi, was analysed. Fig.~\ref{fig4} shows the experimental results which gave anisotropies up to about 7\% at the lowest temperatures. The raw data were taken over a period of about 5 days and all decay corrections were made using the simple 3.66 d half-life of  $^{\rm 224}$Ra. The observed temperature dependence was fitted with the magnetic dipole moment of  $^{\rm 212}$Bi as the unknown quantity, yielding the value $\mid\mu\mid$ = 0.41(5) n.m. \cite{lin92b}, which is in very reasonable agreement with a subsequent laser-based measurement of 0.32(4) n.m. \cite{kil97}. These observations are in complete contradiction to the Full scenario prediction that the the $^{\rm 212}$Bi activity would follow a 10.6 h lifetime throughout the low temperature experiment.
 
As Fig.~\ref{fig3} shows, on even the Partial scenario, the three sets of 1 K `warm' counts would have been seriously discrepant, the first and third lying close to the time dependence predicted by the Null scenario, but the second being above them by 24\%, which was not found, all three sets being in good statistical agreement. A second, equally unobserved, prediction is that the counting rates in all detectors would have been shifted by 36\% on cooling to millikelvin temperature. We conclude that the published data provide evidence that any half-life change on cooling to 1K is well below 5\% (the Partial scenario assumption), less than 1 part in 10$^{\rm 5}$ of the Full scenario prediction, and that a realistic upper limit of any effect is of order 1\%. The same upper limit of 1\% applies to any changes between 1 K and 25 mK.

\subsection{The $^{\rm 225}$Ra decay chain.}  
\begin{figure}
\vspace*{0.3cm}
\centerline{\psfig{file=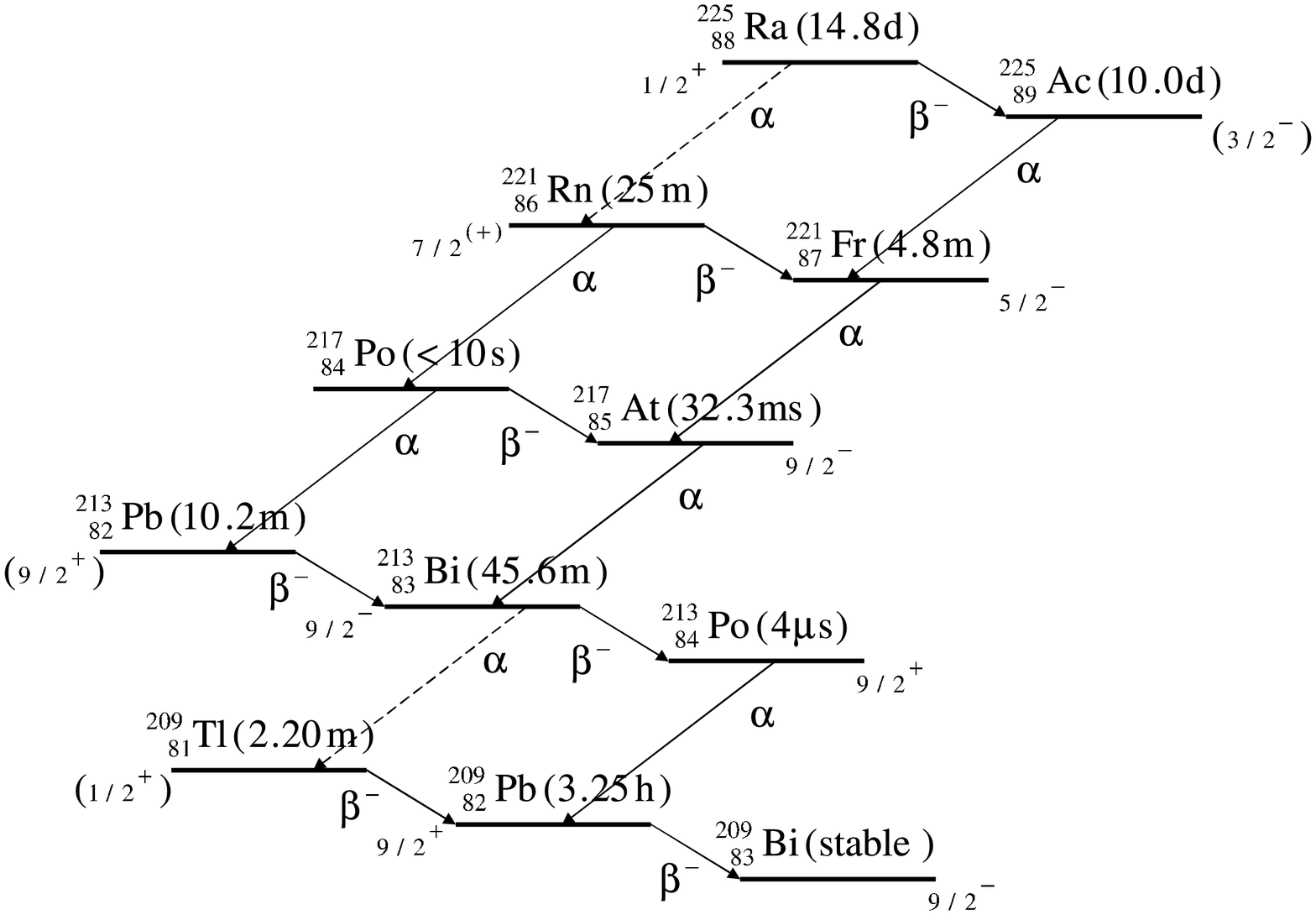,width=10cm}}
\caption{\label{fig5}$^{\rm 225}$Ra decay chain with additional transitions if alpha decay enhancement were strong in  $^{\rm 225}$Ra and $^{\rm 213}$Bi.}
\end{figure}
The case of the $^{\rm 225}$Ra decay chain is basically different since the long-lived parent, $^{\rm 225}$Ra normally decays by beta decay with a half-life of 14.8 d to $^{\rm 225}$Ac with no alpha branch and is not predicted to show (major) half-life change. $^{\rm 225}$Ac, with `normal' half-life 10.0 d then initiates an alpha decay sequence to $^{\rm 213}$Bi. Discussion of possible changes in the alpha spectrum caused by changing $\beta^-$/$\alpha$ branching ratios in this decay chain (e.g. $^{\rm 225}$Ra and $^{\rm 213}$Bi) is omitted as such effects should be much stronger in the $^{\rm 227}$Ac decay (see below). 

\begin{figure}
\vspace*{0.3cm}
\centerline{\psfig{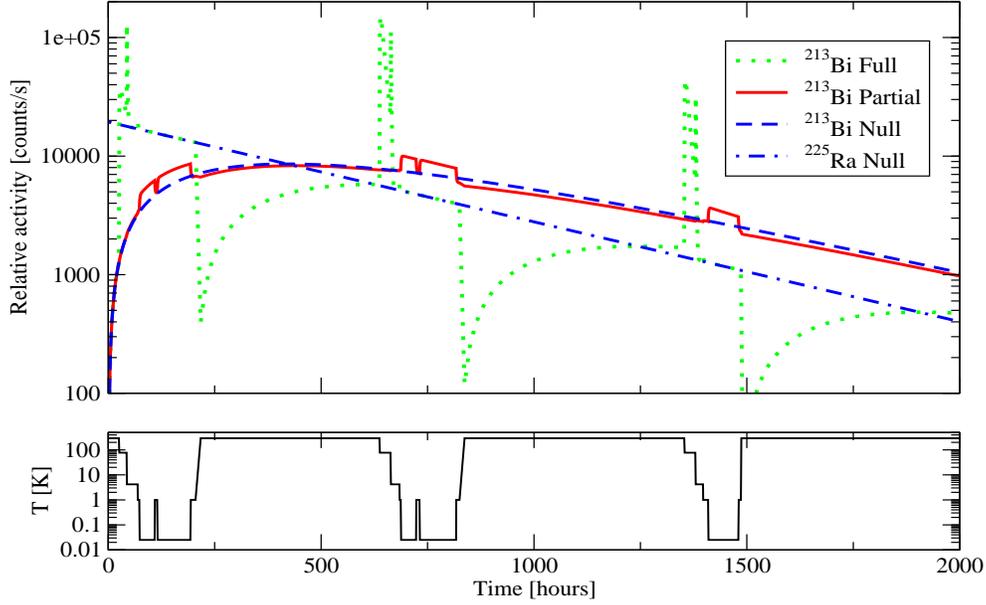}}
\caption{\label{fig6}Relative activity for $^{\rm 213}$Bi and $^{\rm 225}$Ra in the $^{\rm 225}$Ra decay chain in the Full, Partial and Null scenarios (top panel), and sample temperature (bottom panel) as a function time after sample implantation.}
\end{figure}

The predicted behaviour on cooling under the three scenarios is illustrated in Fig.~\ref{fig6}. As shown in the figure, this sample was the subject of three experiments. The first two took place at ISOLDE, CERN, using the NICOLE dilution refrigerator. Of these the first started about 24 hours after the sample was made and continued for about 350 hours whilst the second started after about 750 hours and ended after about 820 hours. The sample was then taken to Oxford and the last experiment was run with the sample between about 1300 and 1500 hours old. For each experiment the temperature sequence was similar to the $^{\rm 224}$Ra chain experiment described above.
In Fig.~\ref{fig6} the decay of the parent activity is a straight line with `normal' half-life of 14.8 d, whilst the `normal' (10.0 d) decay of $^{\rm 225}$Ac and its much shorter lived daughters should follow the rising and then falling smooth curve, which is the Null scenario prediction for this decay chain.  The Full scenario prediction is that, in each of the three experiments, the $^{\rm 225}$Ac activity, and that of its immediate daughters, should show a large increase to well above the `normal' growth curve at 77 K, which disappears over about 30 hours when the sample is held at this temperature. Before equilibrium is re-established however, the temperature has been reduced to 4.2 K, leading to a second increase in  count rate, which died away within less than hour. The activity then settles into simple equilibrium with the decay of $^{\rm 225}$Ra parent, with a 14.8 d half-life. Subsequent cooling to 1 K and to millikelvin temperature, and return to 1 K during the experiment, has little effect in this scenario since there is very little activity `held-up' in the decay chain for such short daughter half-lives.   

The $^{\rm 213}$Bi beta activity will show similar strong short term increases, which recover with close to its `normal' half-life of 45.6 m before also coming to equilibrium with the 14.8 d decay, showing no effect of the intervening activities. It is predicted to decay with a 14.8 d half-life for the duration of the sub-4.2 K periods of all experiments. 

After each experiment at low temperature, when the sample is returned to room temperature, the activities are predicted to fall dramatically then recover towards the Null scenario curve. This recovery has the time constant of the `normal' 10.0 d $^{\rm 225}$Ac half-life and so should take about three weeks, longer than the interval between the first and second experiments.

In Fig.~\ref{fig6} the Partial scenario is seen as producing relatively minor, but still distinctly visible, changes compared to the Null scenario. Any effect before reaching 1 K is small, but the Partial scenario gives a predicted activity increase of 28\% in all detectors on cooling between 1 K and 25 mK.

\subsubsection{Experimental findings for the $^{\rm 225}$Ra decay chain.}
\begin{figure}
\vspace*{0.3cm}
\centerline{\psfig{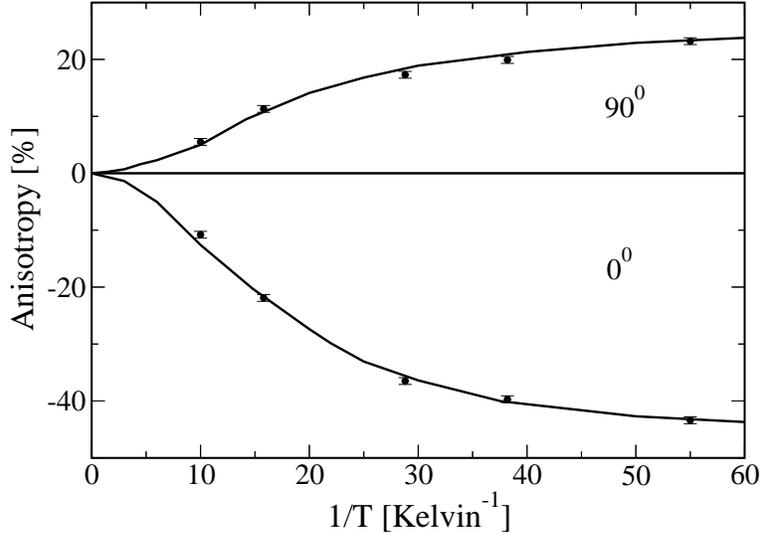}}
\caption{\label{fig7}$^{\rm 225}$Ra decay chain: $^{\rm 213}$Bi 292 keV gamma transition axial (0$^{\rm 0}$) and equatorial (90$^{\rm 0}$) anisotropy vs inverse temperature \protect\cite{lin92a}.}
\end{figure}

\begin{figure}
\vspace*{0.3cm}
\centerline{\psfig{file=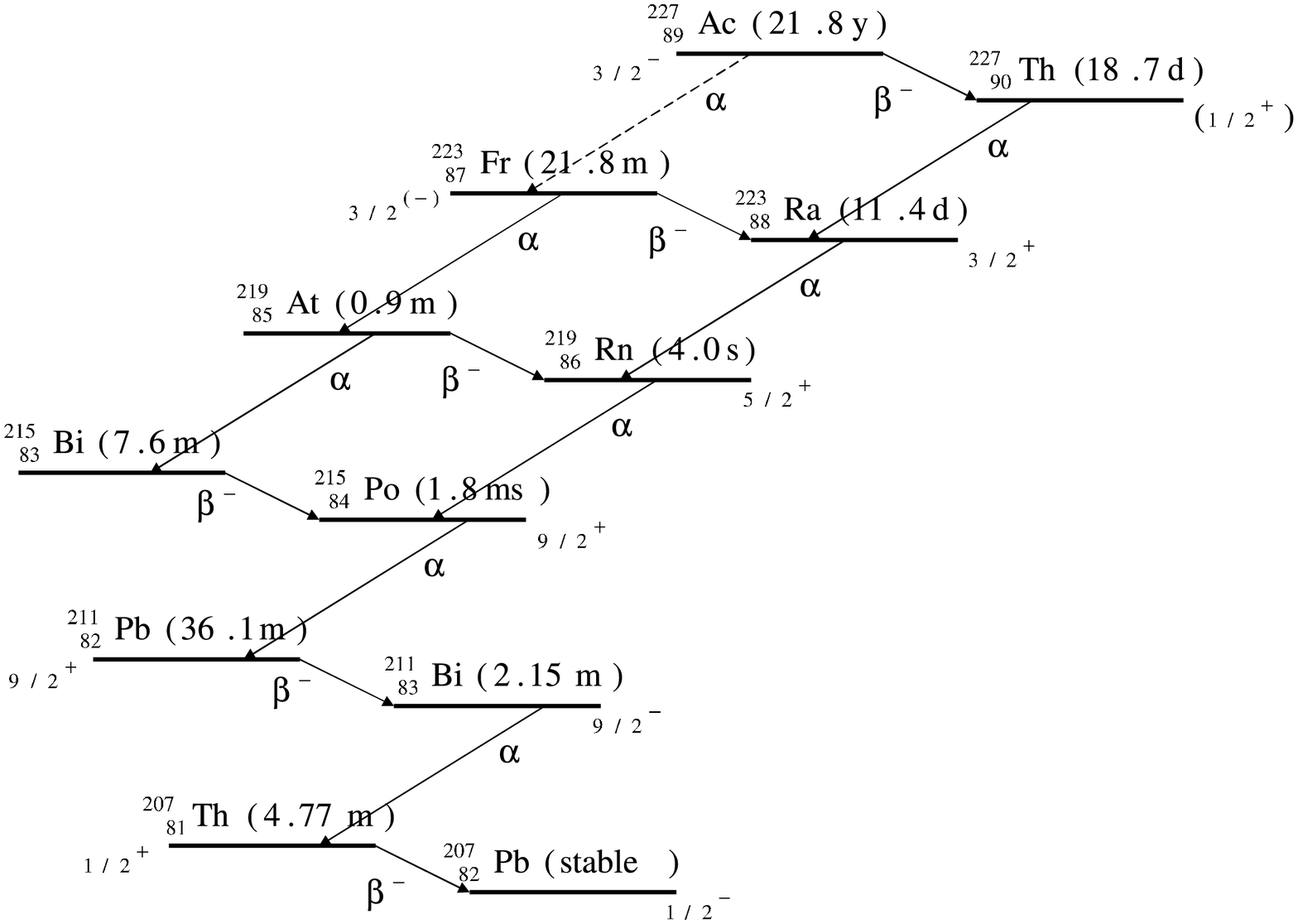,height=8cm,width=10cm}}
\caption{\label{fig8}$^{\rm 227}$Ac decay chain with additional transitions if alpha decay enhancement 
were strong in $^{\rm 227}$Ac.}
\end{figure}

In the first two experiments on this chain only gamma decay data were taken, whilst in the third alpha decay spectra were also observed down to millikelvin temperatures. The published record shows anisotropies measured on the alpha transitions from $^{\rm 225}$Ac, on both alpha and gamma transitions from $^{\rm 221}$Fr and from the alpha decay of $^{\rm 217}$At, all precursors of the decay of $^{\rm 213}$Bi \cite{lin92b}. Data for the 292 keV gamma transition in the decay of $^{\rm 213}$Bi is shown in Fig.~\ref{fig7}, which gives results for detectors along (0$^{\rm 0}$) and normal to (90$^{\rm 0}$) to the axis of nuclear orientation. The data throughout the decay sequence were treated with decay corrections based on the `normal' lifetimes of the longer lived components of the chain,$^{\rm 225}$Ra (14.8 d) and $^{\rm 225}$Ac (10.0 d), following the Null prediction.
Several features of all three experiments give clear evidence contrary to the Full scenario. The most obvious is that the time dependence of the 1 K `warm' counts was in agreement, within small $\sim$1\% statistical variation, with the Null prediction, in all experiments even though the time variation during each experiment differed depending upon the age of the sample (see Fig.~\ref{fig6}). Secondly, very different temperature dependence was shown for the alpha anisotopies in the decays of $^{\rm 221}$Fr (`normal' T$_{\rm 1/2}$= 4.8 m) and $^{\rm 217}$At (`normal' T$_{\rm 1/2}$= 32 ms). This is only possible, as outlined in Sec.~\ref{sec:ltno}, if there is relaxation of these nuclei to achieve thermal equilibrium with the surrounding metallic iron lattice within their lifetimes. In the Full scenario, from Table.~\ref{tab1}, the lifetimes of $^{\rm 221}$Fr and $^{\rm 217}$At become 34 ms and 18 $\mu$s at 1 K. Both become less than 1 $\mu$s at 25 mK, far shorter than the relatively well known Korringa mechanism would be able to relax to equilibrium \cite{sha89}. Thus these activities should not become appreciably oriented unless fed from an already oriented source. Thus  $^{\rm 221}$Fr and $^{\rm 217}$At should both show inherited from the parent $^{\rm 225}$Ac. However, $^{\rm 225}$Ac has spin 1/2 and for spin 1/2 isotopes no anisotropy is observed in emissions which conserve parity. Thus no orientation detected by alpha or gamma decay should be observed in emissions from either $^{\rm 221}$Fr or $^{\rm 217}$At in the Full scenario in direct contrast with experiment \cite{lin92b}.

To place an upper limit on any possible effect, predictions of the Null and Partial scenarios are again compared (Fig.~\ref{fig7}). The most important feature of the Partial scenario is that there is a clear discontinuity in activity as the sample is cooled to 25 mK from 1K which, as detailed above, must distort the 0$^{\rm 0}$/90$^{\rm 0}$ anisotropy relationship. The data on the 292 kev transition are of high statistical accuracy ($\sim$0.5\%) and show excellent agreement with theory, having the correct ratio to one another and both tending to a constant value as the temperature approaches 16 mK (1/T = 60 K$^{\rm -1}$). Any deviation greater that 2\% would be revealed, as compared with the predicted shifts (which are both upward, i.e to increase the 90$^{\rm 0}$ effect and reduce that at 0$^{\rm 0}$) of $\sim$25\% in the Partial scenario. 

The observed anistropies show no sign of such distortions. Both alpha and gamma results were analysed to give the magnetic dipole moment of $^{\rm 213}$Bi. The value obtained, $\mid\mu \mid$ = 3.89(9) n.m. \cite{lin92b}, is in very good agreement with a subsequent laser-based measurement of 3.716(7) n.m. \cite{kil97}.

On this basis the evidence from the $^{225}$Ra mass chain is that there is no half-life effect above 0.5\%.

\subsection{The $^{\rm 227}$Ac decay chain.}  
\begin{figure}
\vspace*{0.3cm}
\centerline{\psfig{file=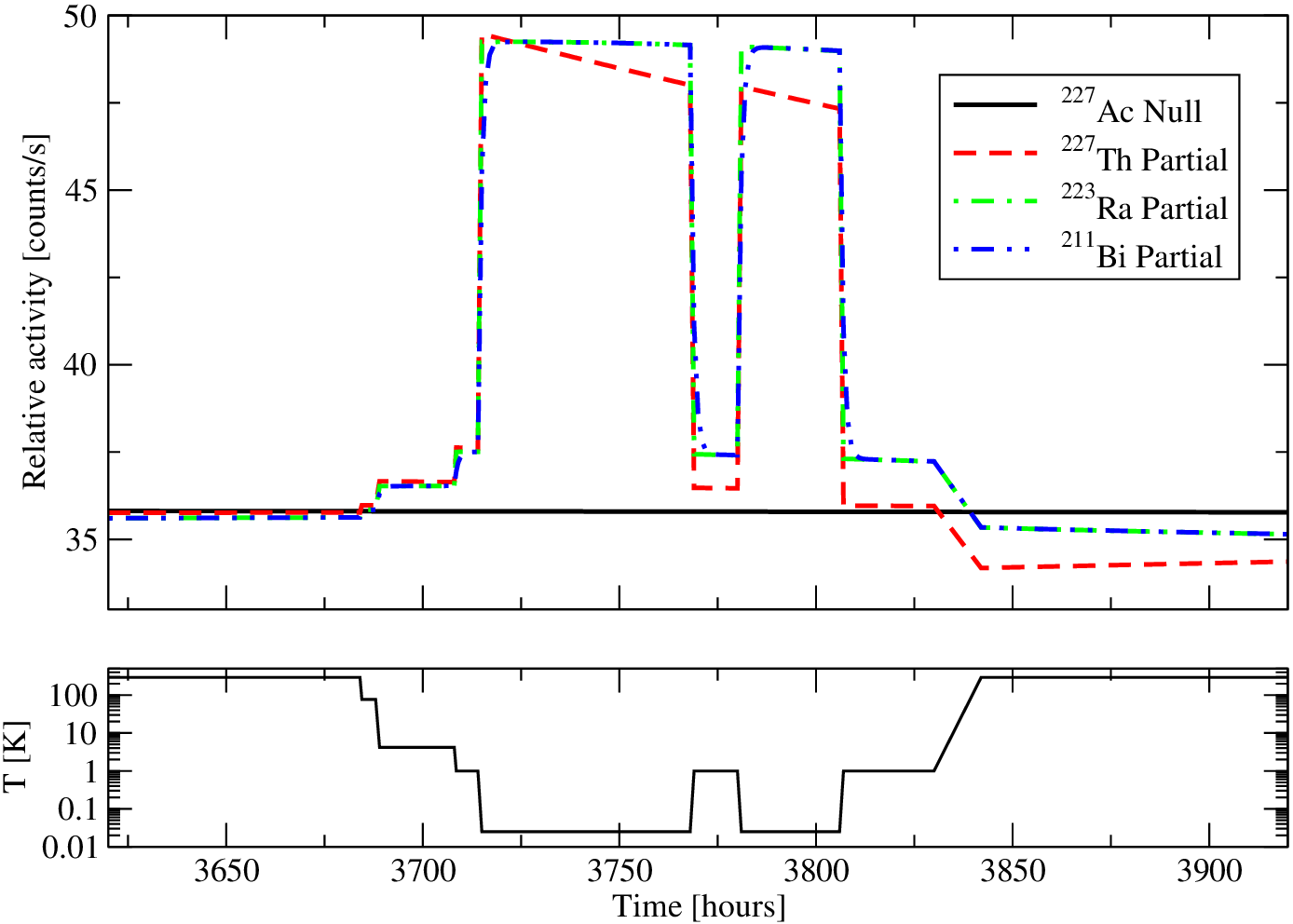,height=8cm,width=13cm}}
\caption{\label{fig9}Relative activity for selected members of the $^{\rm 227}$Ac decay chain in the Null and Partial scenarios (top panel), and sample temperature (bottom panel) as a function of time after implantion.}
\end{figure}

The $^{\rm 227}$Ac chain is shown in Fig.~\ref{fig7}, including predicted changes associated with the Full scenario described below. The `normal' sequence is by dominant (98.6\%) beta decay to $^{\rm 227}$Th, with half-life 21.8 y, followed by the strong alpha sequence $^{\rm 227}$Th -- $^{\rm 223}$Ra -- $^{\rm 219}$Rn -- $^{\rm 215}$Po -- $^{\rm 211}$Pb. There is a weak (1.4\%) alpha decay branch to $^{\rm 223}$Fr, which returns to the main sequence through (99\%) beta decay of $^{\rm 223}$Fr to $^{\rm 223}$Ra. A second weak (0.005\%) alpha branch ($E_\alpha \sim$ 5300 keV) exists from $^{\rm 223}$Fr to $^{\rm 219}$At which then decays further by dominant alpha decay to $^{\rm 215}$Bi.
The Null scenario prediction is that this decay follows a time dependence basically determined by the time for the two longer-lived alpha decays, $^{\rm 227}$Th (18.7 d) and $^{\rm 223}$Ra (11.4 d), to come into temporal equilibrium with the 21.8 y $^{\rm 227}$Ac activity. Fig.~\ref{fig9} shows that this was complete before the experiment, which took place approximately 150 days after the sample was implanted. Both alpha and gamma transitions from all elements of the decay chain were observed and alpha spectra are shown at room temperature (Fig.~\ref{fig10}) and at 1 K and $\sim$10 mK (Fig.~\ref{fig11}).

 The Full scenario estimated half-life changes from Table~\ref{tab1} alter this sequence fundamentally at 1 K ( and there will be less striking changes at higher temperatures). Through alpha decay enhancement, the $\beta^-$/$\alpha$ branching ratio  for $^{\rm 227}$Ac is altered from 70 to 1.5 x 10$^{\rm -4}$ so $\alpha$ decay to $^{\rm 223}$Fr should become dominant. Furthermore the subsequent decay of $^{\rm 223}$Fr also becomes dominated by alpha decay (86\%) with beta decay reduced to 14\%. Thus, as the sample is cooled, not only should the total activity show marked increases but very strong feeding of $^{\rm 219}$At and $^{\rm 215}$Bi should be observed before the source loses activity with a half life of about 1 day, rather than the normal 21.8 y.
A further major prediction of the Full scenario for this decay chain is that the alpha decay fine structure observed in the decays of $^{\rm 227}$Th and $^{\rm 219}$Rn will be seriously altered, the higher energy alpha transitions becoming weaker relative to the lower energy transitions. The calculations presented in Table~\ref{tab1} predict a doubling at 1 K of the intensity of the 5756 keV transition in the decay of $^{\rm 227}$Th relative to the 6038 keV transition. In $^{\rm 223}$Ra the relative intensity of the 5434 keV alpha branch should increase from 2.3\% to 4.4\% whilst the 5717 keV branch is predicted to weaken from 54\% to 47\%. These combined changes would seriously modify the appearance of the spectral region between 5300 and 6100 keV in the spectrum from this sample at 1 K as compared to room temperature. Cooling to millikelvin temperatures would, in this scenario, lead to further major changes in half-life and fine structure relative intensity. 

The Partial scenario predictions for the experiment performed on the $^{\rm 227}$Ac decay chain are shown in Fig~\ref{fig9}. The small change in the $^{\rm 227}$Ac  $\beta^-$/$\alpha$ branching ratio has been neglected so that the experiment starts close to the temporal equilibrium situation of the Null scenario. Cooling to 4.2 K produces $\sim$1.6\% increase in all activities, with a further 1.2\% increase at 1 K and a much larger ($\sim$30\%) increase on cooling to 25 mK. These increased activities die away only slowly during the 1 K and 25 mK sections of the experiment, with the exception of  $^{\rm 227}$Th which shows a decay with the reduced lifetime of $\sim$13 d throughout in this period. This decay is far faster than the Null scenario and has direct consequence for the sequence of 1 K `warm' counts on alpha transitions from this isotope.
\subsubsection{Experimental findings for the $^{\rm 227}$Ac decay chain}
\begin{figure}
\vspace*{0.3cm}
\centerline{\psfig{file=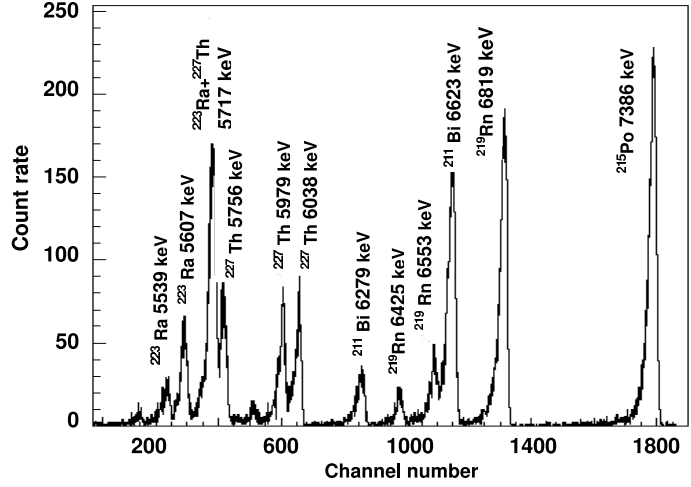,width=10cm}}
\caption{\label{fig10}Room temperature spectrum of $^{\rm 227}$Ac chain alpha decay.}
\end{figure}
\begin{figure}
\vspace*{0.3cm}
\centerline{\psfig{file=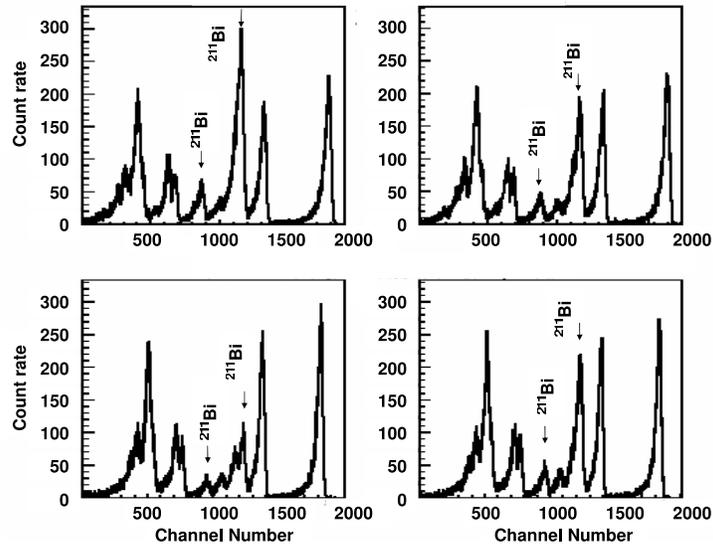,width=10cm}}
\caption{\label{fig11}$^{\rm 227}$Ac decay chain: Alpha spectra in two detectors, upper panels at 86$^{\rm 0}$, lower at 24$^{\rm 0}$, to the polarization axis. The right-hand panels show spectra taken at 1 K, left-hand panels at $\sim$10 mK.}
\end{figure}
\begin{figure}
\vspace*{0.3cm}
\centerline{\psfig{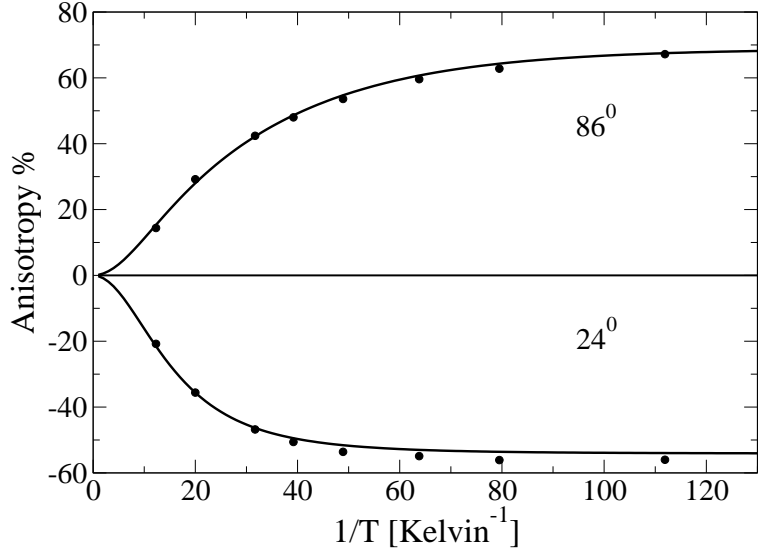}}
\caption{\label{fig12}$^{\rm 227}$Ac decay chain: $^{\rm 211}$Bi 6623 keV alpha transition anisotropy vs inverse temperature \protect\cite{wil96}.}
\end{figure}

In the data analysis, all decay corrections were made with the 21.8 y $^{\rm 227}$Ac half-life, the Null scenario, and are very close to unity. The experimental evidence that the $^{\rm 227}$Ac decay chain shows little or no half-life change on cooling has many aspects of which three are considered. The 1 K and $\sim$10 mK spectra in Fig.~\ref{fig11} were taken for equal times. There is clearly no overall change in activity on cooling from 1 K to the lowest temperature, although anisotropy in the $^{\rm 211}$Bi transitions is readily seen. \\ 
\noindent
(i) The close similarity of the alpha spectra at room temperature, 1 K and 10 mK, Figs.~\ref{fig10},~\ref{fig11} shows that the predictions of the Full scenario introducing new alpha transitions and eliminating others, did not occur. The  $^{\rm 227}$Th,  $^{\rm 223}$Ra and  $^{\rm 219}$Rn peaks were predicted to vanish before reaching 1 K.\\
\noindent
(ii) The observed lack of anisotropy in decay of $^{\rm 227}$Th (spin 1/2), $^{\rm 223}$Ra (weak hyperfine interaction) and $^{\rm 215}$Po (extremely short half-life compared to relaxation time) are all in agreement with predictions of theory for the reasons cited \cite{Wthesis}. These results would be strongly distorted in the Partial scenario by the 30\% increase in activity on cooling from 1 K to 25 mK.\\
\noindent
(iii) The observed large anisotropies of the 6623 keV alpha transition from  $^{\rm 211}$Bi (see Fig.~\ref{fig12}) have small experimental errors. They show the correct relationship in two directions, 86$^{\rm 0}$ and 24$^{\rm 0}$, one requiring increase and the other decrease of the recorded activity, and strong temperature dependence leading to constant maximum values at the lowest temperatures reached. As discussed above for the  $^{\rm 213}$Bi 292 keV transition in the $^{\rm 225}$Ra chain, all these aspects of the data would be disturbed by any increase in activity between 1 K and the lowest temperatures caused by changing lifetime. For this case the precision of each data point in Fig.~\ref{fig12} is $\sim$1\%. The data were fitted to give the magnetic dipole moment of  $^{\rm 211}$Bi yielding the value $\mid\mu\mid$ = 3.79(7) n.m. in excellent agreement with theoretical prediction \cite{wil96,blo65,huy73}. The predicted upward shift of all data in this figure in the Partial scenario is close to 30\% and would be impossible to miss.
Thus the limit on half-life change from the $^{\rm 227}$Ac mass chain is 1\% or lower, based on the lack of activity increase in cooling from 1K to $\sim$25 mK, on the constancy of (normally corrected) warm counts and detailed evaluation of data on many transitions. The absence of change in the alpha spectrum, although well established qualitatively, places a less stringent limit on possible half-life change since to be clearly observed it requires considerable increase in the alpha decay branching from  $^{\rm 227}$Ac.
\section{\label{sec:discon}Discussion and conclusions}
In this paper evidence has been presented concerning the prediction that introducing alpha activities into metals and cooling them to temperatures of a few Kelvin and below should produce major changes in their decay rates and the appearance of their alpha spectra. On this evidence, no such changes have been observed with an upper level of one percent, at 1 K, a factor of about 10$^{\rm 5}$ smaller than the predicted changes. Two of the three nuclear moments deduced from these experiments, which depended upon correct source decay treatment, have been confirmed by independent subsequent measurements; the third, measured solely by this technique, agrees well with theoretical calculation. 

The evidence presented has been of five types, namely
\begin{enumerate}
\item consistency of warm counts corrected using 'normal' half-lives - a 1 K property.
\item absence of activity change between 1 K and millikelvin temperatures as shown by
smooth temperature variation of observed anisotropies and the correct low temperature saturation behaviour of strong anisotropies. Large effects are predicted, even in the Partial scenario. However, these predictons depend upon accepting the T$^{\rm -1/2}$ dependence of the screening energy U$_{\rm D}$ as applicable down to millikelvin temperature - the original suggestions in Ref.~\cite{ket06} made no limit of temperature range and took as their example cooling to 4.2 K. 
\item absence of anisotropy where isotope properties lead to no polarisation.
 \item no loss of peaks or appearance of additional peaks in alpha spectra between room temperature and millikelvin temperatures.
\item evidence that ratios of relaxation time to half-life are not changed enough to prevent thermal equilibrium in some shorter lived ground states.
\end{enumerate}
 Types 3--5 are qualitative rather than quantitative features which, whilst denying the possibility of the Full scenario are more difficult to use to place specific limits on much smaller effects.

The emphasis in this paper has been on predictions for alpha decay, not only because the effects suggested are so large, but also because, if confirmed they would have important consequences with regard to long-lived waste disposal. However the same ideas also predict more modest changes in beta decay half lives \cite{lim06}. The LTNO technique has, for more than forty years, studied radioactive samples cooled to sub-Kelvin temperatures. Since the introduction of methods of cooling of metal samples to millikelvin temperatures in the early 1960's, with activity melted, diffused and, more recently, implanted into  metals, it has been a routine element of such experiments to correct for decay in the source activity. Such corrections have always used the `normal' half-life of the isotopes concerned. Many tens of different beta-emitting isotopes have been oriented, with lifetimes ranging from many years to a few hours and even less. The host metals used in different studies include the more common iron, cobalt and nickel but also copper, silver, gold and many lanthanides. No evidence for any deviation from the `normal' lifetime for these isotopes has been reported by the many research groups involved in this work (for comprehensive listing see \cite{sto86} and references therein or search for results on low temperature nuclear orientation). Although not directly addressed in this paper, a firm upper limit on any lifetime changes associated with cooling beta activities in pure metals can be set at or below 1\%. (The well established phenomenon of change in electron capture half-lives with variation of inner electron concentration in different environments is excluded from this remark.)

Since Ref.~\cite{ket06} appeared there have been several follow-up papers. Amongst these the original proponents of the effect, Raiola et al. \cite{rai07} claim to have observed a 6\% lifetime decrease on cooling a sample of $^{\rm 210}$Po, implanted into Cu, to 12K -- a result `consistent in sign but significantly smaller than expected' \cite{rai07} ~(i.e. 6\% compared to 1000\%, or 1 part in 150 of the prediction). Jeppesen et al. \cite{jep07} report changes in half-life at room temperature between decay of $^{\rm 221}$Fr implanted into metals and non-metals of order 0.4\%, with comparable errors, where the Full prediction would be reduction by a factor $\sim$2 (1 part in 500 of the prediction).

The analysis presented in this paper places much more stringent limits on the size of any effect. The upper limit of 1\% on any effect at 1 K, compared to predictions of factors above 10$^{\rm 4}$ represents 1 part in 10$^{\rm 6}$. The evidence for lack of effect down to 25 mK would, if the proposed theory were simply accepted at all temperatures, place limits lower than 1 part in 10$^{\rm 10}$.  

One may conclude that the suggestion of alleviating problems associated with the disposal of long-lived radioactive waste by cooling in pure metals is unrealistic. Concerns expressed in Ref.~\cite{ket06} regarding the reliability of other less topical but scientifically equally important situations where reliance is placed upon the use of `normal' half-lives in a metallic environment, are also misplaced.
\section{Acknowledgement}
The authors acknowledge support from the NICOLE and ISOLDE (CERN) Collaborations with preparation of the radioactive samples and assistance with some of the reported experiments.
This work was supported by grants from EPSRC (UK), and by US DOE grants No. DE-FG02-96ER40983 (UT) and No. DE-FG02-94ER40834 (UMD).

\end{document}